\documentclass{PoS}

\newcommand{\ee}{e^{+}e^{-}}

\newcommand{\uubar}{u\bar{u}}
\newcommand{\ddbar}{d\bar{d}}
\newcommand{\ssbar}{s\bar{s}}
\newcommand{\qqbar}{q\bar{q}}
\newcommand{\ccbar}{c\bar{c}}

\newcommand{\jp}{J/\psi}

\newcommand{\psip}{\psi(2S)}

\newcommand{\mumu}{\mu^{+}\mu^{-}}
\newcommand{\pipi}{\pi^{+}\pi^{-}}

\newcommand{\pim}{\pi^{-}}

\newcommand{\bbar}{B\bar{B}}

\newcommand{\rt}{\rightarrow}

\newcommand{\etal}{\em et al.}

\newcommand{\yones}{\Upsilon(1S)}
\newcommand{\ytwos}{\Upsilon(2S)}
\newcommand{\yonetwos}{\Upsilon(1S,2S)}
\newcommand{\yns}{\Upsilon(nS)}

\newcommand{\lm}{\Lambda}

\newcommand{\lmb}{\bar{\Lambda}}
\newcommand{\lmlm}{\Lambda\Lambda}

\newcommand{\lmppi}{\Lambda p \pi^-}
\newcommand{\lmbpbpi}{\bar{\Lambda} \bar{p} \pi^+}

\newcommand{\xim}{\Xi^{-}}

\usepackage{wrapfig}

\title{Properties of non-$q\bar{q}$ $XYZ$ mesons and results of a search for the $H$-dibaryon}

\ShortTitle{$XYZ$ mesons and $H$-dibaryon search}

\author{\speaker{Stephen Lars Olsen}\\
        Department of Physics \& Astronomy, Seoul National University\\
        Gwanak-gu, Seoul, 151-747, KOREA\\
        E-mail: \email{solsen@hep1.snu.ac.kr}}

\abstract{A number of charmonium- and bottomonium-like meson states have been
observed that have properties that do not match well to expectations
for the simple quark-antiquark substructure suggested by the constituent
quark model.  Some of them are electrically charged and decay to final
states containing hidden charmonum or bottomonium mesons and, thus, must 
contain at least four quarks.  Common properties of these so-called 
$XYZ$ mesons are partial widths for decays to hidden quarkonium
states plus light hadrons that are much larger than corresponding
partial widths for established quarkonium mesons. I review some
recent results from the Belle experiment, including the
recent discovery of two charged bottomium-like states, the
$Z_b(10610)^\pm$ and $Z_b(10650)^\pm$, that
decay to $\pi^\pm h_b(mS)$ ($m=1,2$) and 
$\pi^\pm \Upsilon(nS)$ ($n=1,2,3$) final states. 
In addition, I present recent Belle results from a search for $H$-dibaryon 
production in inclusive $\Upsilon(1S)$ and $\ytwos$ decays.
}

\FullConference{International Winter Meeting on Nuclear Physics,\\
		21-25 January 2013\\
		Bormio, Italy }

\begin{document}

\section{Introduction}
\noindent
According to the prescriptions of the original quark model proposed by
Gell-Mann~\cite{gellmann} and Zweig~\cite{zweig}
in 1964,  mesons are comprised of quark-antiquark pairs and baryons are
three-quark triplets.  In the 1970's, this simple model was superseded by
Quantum Chromodynamics (QCD), which identified the reason for these rules
was that $\qqbar$ pairs and $qqq$ combinations can be color singlet
representations of the color $SU(3)$ group that is fundamental to the theory.
Somewhat suprisingly, the mesons are $\qqbar$ and baryons are $qqq$
prescription still adequately describes the hadronic particle spectrum
despite the existence of a number of other color-singlet quark and gluon
combinations that are possible in QCD~\cite{strottman}.
Considerable experimental efforts
at searching for the predicted color-singlet $qqq\bar{q}q$ ``pentaquark''
baryons~\cite{diakonov}
and the doubly strange $udsuds$ $H$-dibaryon~\cite{jaffe_H} have failed to come up with
any unambiguous candidates for either state~\cite{trilling}.
Although a few candidates for non-$\qqbar$ light hadron resonances
have been reported~\cite{e852} none have been generally accepted as
established by the hadron physics community~\cite{barnes_1}.

In recent years, however, the situation changed, beginning with the observation of
the $X(3872)$ meson by Belle~\cite{skchoi_x3872}, the
discovery of the $Y(4260)$ meson by BaBar~\cite{babar_y4260}, and the
subsequent observation of a number of  other candidate charmonium-like meson states, 
the so-called $XYZ$ mesons, that are not well matched to expectations for the
quark-antiquark meson picture~\cite{godfrey_olsen}.  Here I give a brief report
on why we think the observed states may be exotic and describe some recent
observations of charged quarkonium-like meson states that
necessarily must have a minimal four-quark structure
by Belle~\cite{skchoi_z4430, mizuk_z4050,bondar_zb}
and BESIII~\cite{bes_z3900}. 

In 1977, Jaffe
predicted the existence of
The $H$-dibaryon, a doubly strange,
six-quark structure ($uuddss$) with quantum numbers
$I=0$ and $J^P=0^+$ and a mass that is $\simeq 80$~MeV
below the $2m_{\lm}$~\cite{jaffe_H}.   An $S=-2$, baryon-number $B=2$
particle with mass below $2m_{\lm}$ would decay via weak
interactions and, thus, be long-lived with a lifetime comparable
to that of the $\lm$ and negligible natural width.  

Jaffe's specific prediction was ruled out by the observation
of double-$\lm$ hypernuclei events~\cite{double-lambda,nagara,e176},  
especially the famous ``Nagara'' event that has a relatively unambiguous
signature as a $_{\lmlm}^6$He hypernucleus produced via
$\xim$ capture in emulsion~\cite{nagara}.
The measured $\lmlm$ binding energy,
$B_{\lmlm}=7.13\pm 0.87$~MeV, establishes, with a 90\% confidence level
(CL), a lower limit of $M_{H}>2223.7$~MeV, severely narrowing the window
for a stable $H$ to the binding energy range
$B_H\equiv 2m_{\Lambda}-M_H < 7.9$~MeV\footnote{In this report I have 
taken the liberty of averaging asymmetric errors
and combining statistical and systematic errors
in quadrature.  For actual measured values, please refer
to the original papers.} 

Although Jaffe's original prediction for a binding energy of $\simeq 81$ MeV
has been ruled out,  the theoretical case for
an $H$-dibaryon with a mass near $2m_{\lm}$ continues to be
strong and has been recently strengthened by lattice QCD
calculations (LQCD) by the NPLQCD~\cite{NPLQCD,NPLQCD_2} and
HALQCD~\cite{HALQCD} collaborations that both find 
a bound $H$-dibaryon, albeit for non-physical values for
the pion mass.  
NPLQCD's linear (quadratic) extrapolation to
the physical pion mass gives $B_H= -0.2\pm 8.0$~MeV
($7.4\pm 6.2$~MeV)~\cite{NPLQCD_2}.  Carames and
Valcarce~\cite{Carames} recently studied the $H$ with a
chiral constituent model constrained by $\Lambda N$,
$\Sigma N$, $\Xi N$ and $\lm\lm$ cross section data
and find  $B_H$ values that are similar to the NPLQCD
extrapolated values. 

Numerous experimental searches have been made for an
$H$-dibaryon-like state with mass near (above or below)
the $2m_{\lm}$ threshold.  Although some hints
of a virtual $\lm\lm$ state was reported by a
KEK experiment~\cite{e522}, other searches produced
negative results~\cite{e836,KTeV,e910,H-searches}. 

\section{The Quarkonium Spectra}
\noindent
Quarkonium mesons, {\it i.e.,} mesons that contain a $Q$ and $\bar{Q}$ quark pair,
where $Q$ is used to denote either the $c$- or $b$-quark, have proven
to be useful probes for multiquark meson systems.  That is because these mesons are well understood;
their constituent quarks are non-relativistic and potential models can be applied.  Most of the low-lying
$Q\bar{Q}$ meson states have been discovered and found to have properties that agree
reasonably well with potential model predictions.  More complex states would likely have properties that
deviate from model predictions and, thus, be identifiable as such.

\begin{figure}[t]
\begin{minipage}{70mm}
\centering
 \includegraphics[height=0.9\textwidth,width=1.1\textwidth]{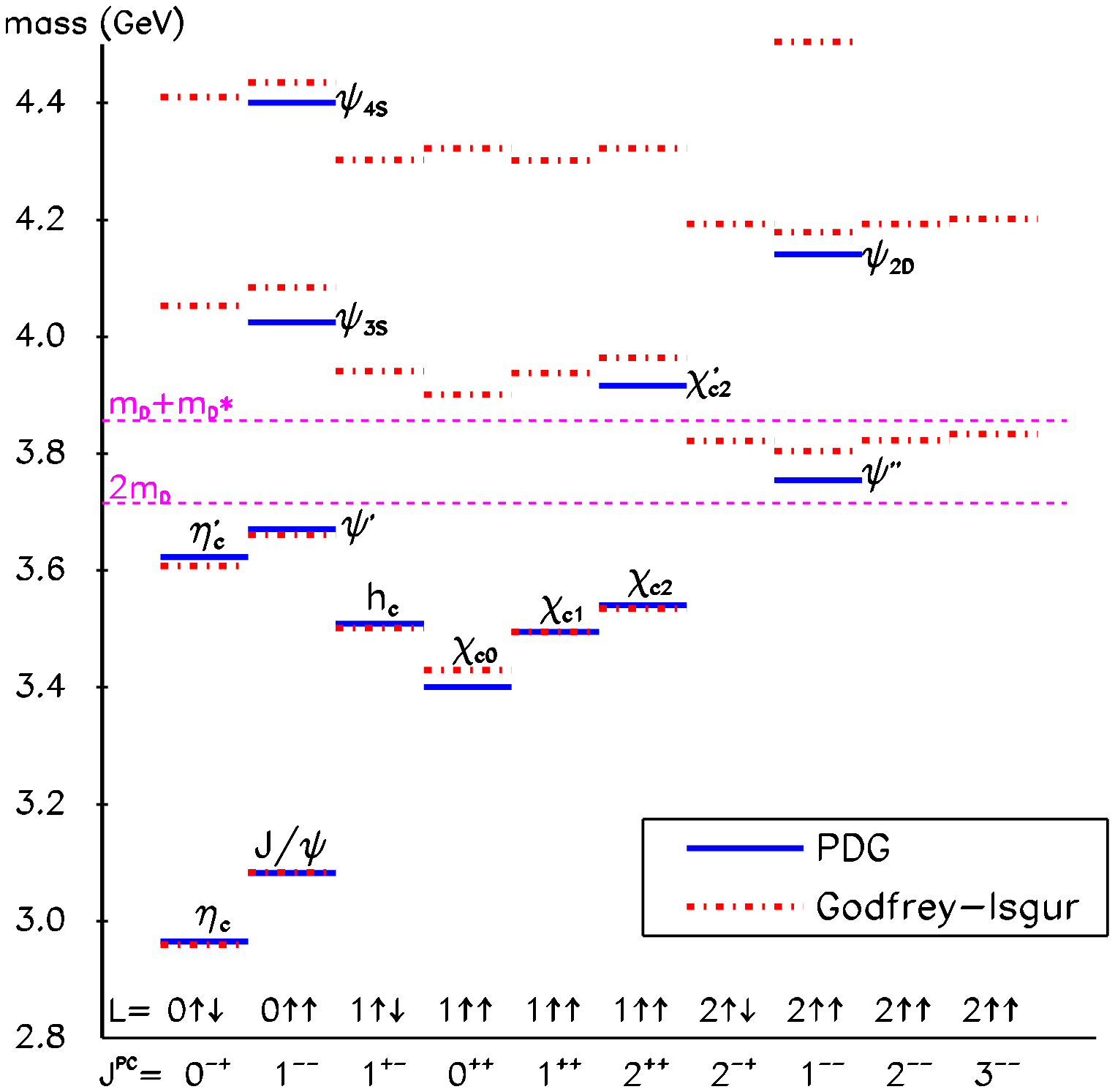}
\caption{ The charmonium meson spectrum.  The solid bars indicate the established
charmonium states and the dash-dot bars indicate the mass levels that were predicted
in 1985.}
\label{fig:charmonium}
\end{minipage}
\hspace{\fill}
\begin{minipage}{70mm}
\centering
  \includegraphics[height=0.9\textwidth,width=1.1\textwidth]{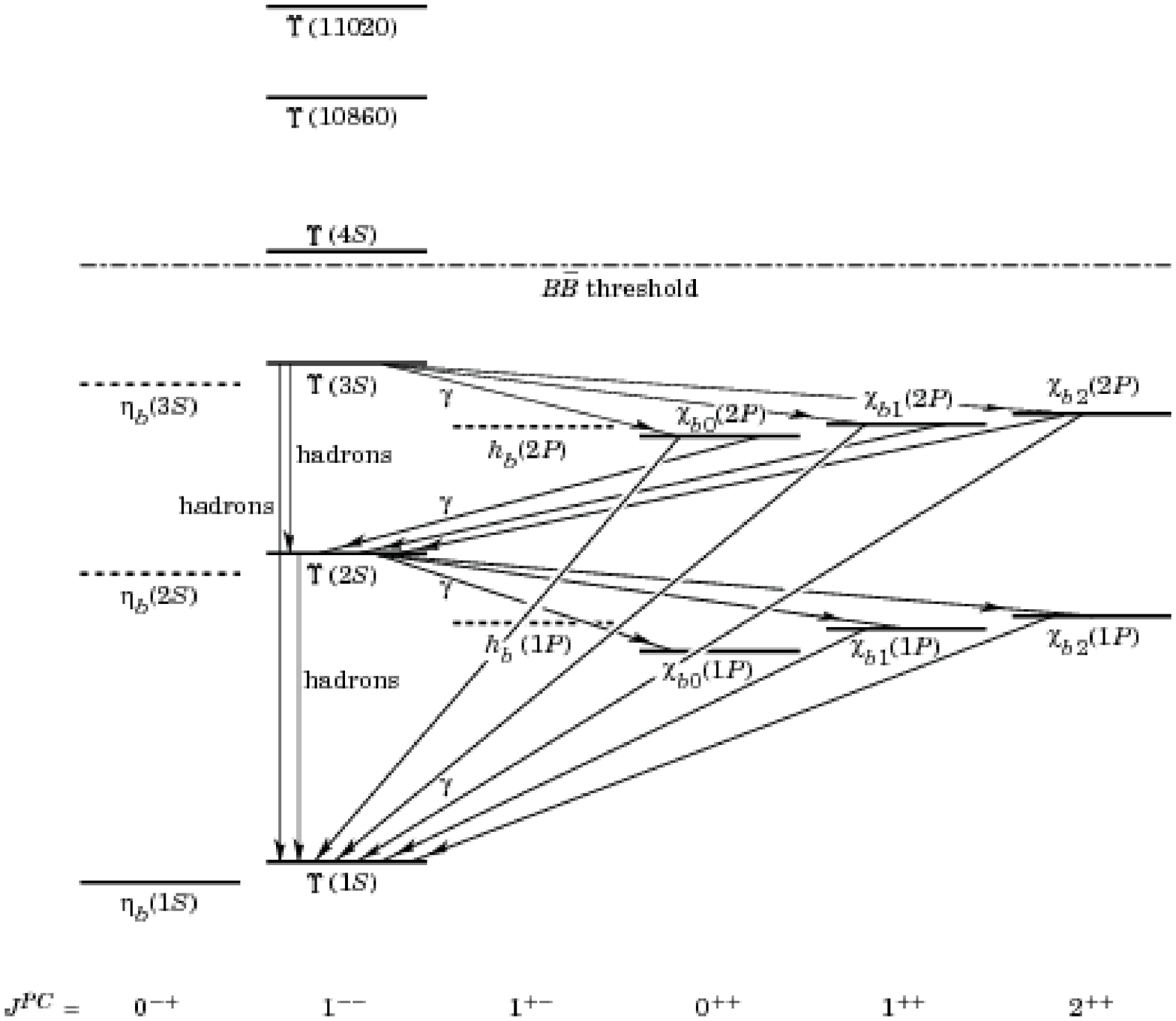}
\caption{The bottomonium level diagram. The solid bars indicate well established
states, in addition, candidates for the $h_b(1P)$, $h_b(2P)$ and $\eta_b(2S)$
states have recently been seen. }
\label{fig:bottomonium}
\end{minipage}
\end{figure}

Figure~\ref{fig:charmonium} shows a level diagram for the $c\bar{c}$ (``charmonium'') system,
where established states are indicated by solid lines, and the masses predicted by the
Godfrey-Isgur (GI) relativized potential model in 1985~\cite{GI} are shown as dash-dot lines.
All of the states below the $M=2m_D$ open-charmed-meson threshold have been identified and
have masses that agree reasonably well with GI predictions.  Moreover, all of
the above-threshold $J^{PC}=1^{--}$ states below $M\simeq 4.45$~GeV have been assigned and, 
here too, there is reasonable agreement with predicted masses.  In addition to the $1^{--}$
states, the $\chi_{c2}^{\prime}$, the $2^{++}$ radially excited $2\ ^3P_2$ state,
has been assigned~\cite{uehara_chi3929} and Belle recently reported strong evidence
for the $\psi_2$, the $2^{--}$ $1^{3}D_{2}$ state~\cite{belle_psi2}.
Any meson state with prominent decays to final states containg a $c$- and
a $\bar{c}$-quark, that does not fit into one of the remaining unassigned $\ccbar$ states
has to be considered exotic.\footnote{
The large value of the $c$-quark mass precludes any substantial production
of $\ccbar$ pairs via fragmentation processes.}

Figure~\ref{fig:bottomonium} shows a level diagram for the $\bbar$ (``bottomonium'')
system.  Here all the levels indicated by solid bars are well established.  In
addition, the have been recently reported of the $h_b(1p)$, $h_b(2P)$ and
$\eta_b(2S)$~\cite{belle_hb,belle_etab2s}, as well as evidence for some
of the (unshown) $D$-wave and $\chi_b (3P)$ states~\cite{dwaves}.  Three
$1^{--}$ states above the $2m_B$ open-bottom threshold have been tentatively
identified as the $\Upsilon(4S)$, $\Upsilon(5S)$ and $\Upsilon(6S)$, and
these are their commonly used names.  The arrows in Fig.~\ref{fig:bottomonium}
indicate transitions between the sates accompanied by either light-hadron
emission (vertical arrows) and $E1$ electromagnetic transitions (diagonal
arrows).  Not shown are the $\Upsilon(3S)\rt \gamma \eta_b(1S)$ $M1$-transition
that was used by BaBar to discover the $\eta_b(1S)$~\cite{babar_etab},
the $\Upsilon(5S)\rt\pipi h_b(1P)$
and $\pipi h_b(2P)$ transitions used by Belle to discover the $h_b(1P,2P)$
states~\cite{belle_hb},
or the $h_b(2P)\rt \gamma \eta_b(2S)$ $E1$ transition that led to the
discovery of the $\eta_b(2S)$~\cite{belle_etab2s}.  With the notable exception of the
$\Upsilon(5S)\rt \pipi \Upsilon(1S,2S,3S)$ and $\Upsilon(5S)\rt \pipi h_b(1S,2S)$
transitions, which are anomalously strong and discussed below, all of the other
transitions have measured strengths that are consistent with theoretical expectations.

\section{The $X(3872)$}
\noindent
The first $XYZ$ meson that was observed is the $X(3872)$, which was seen as a
pronounced peak in the $\pipi\jp$ invariant mass spectrum in exclusive
$B^+\rt K^+\pipi\jp$ decays~\cite{skchoi_x3872,conj}. Decays to
$\gamma\jp$~\cite{belle_gammajp,babar_gammajp,belle_gammajp_1}, 
$\pipi\pi^0 \jp$~\cite{belle_gammajp,babar_3pi}, and $D^0\bar{D^{*0}}$~\cite{x3872_ddbar}
have also been seen.  The $\pipi$ invariant mass distribution
in $X(3872)\rt\pipi\jp$ decays is well described by the hypothesis that the pions
originate from $\rho^0\rt\pipi$ decays~\cite{CDF_pipi,belle_jpc}.  A CDF analysis
of angular correlations among final state particles in $X(3872)\rt\pipi\jp$ ruled
out all possible $J^{PC}$ assignments (for $J\le3$) other than $1^{++}$ and
$2^{-+}$~\cite{CDF_jpc}. A Belle analysis of angular correlations in
$B\rt K X(3872)$; $X(3872)\rt\pipi\jp$ decays found good agreement with the
$1^{++}$ hypotheses with no free parameters; for $2^{-+}$ there is one free
complex parameter and a value for this was found that produces acceptable
agreement with the measured data~\cite{belle_jpc}.  Recently, an
comprehensive analysis of the five-dimensional angular correlations
in the $B^+\rt K^+ X(3872)$, $X(3872)\rt\pipi\jp$, $\jp\rt\mumu$ decay
chain conclusively ruled out the $2^{-+}$ assignment and established, once and
for all, that the $J^{PC}$ of the $X(3872)$ is $1^{++}$~\cite{LHCb_jpc}.

The only unassigned $1^{++}$ charmonium level with a predicted mass near 3872~MeV
is the $\chi_{c1}^{\prime}$, the first radial excitation of the $\chi_{c1}$.
The assignment of the $X(3872)$ to this level has some problems.  
First, the mass is too low.
Potential models predict the mass of the $\chi_{c1}^{\prime}$ to be
around 3905~MeV, where this is pegged to the measured mass of the 
multiplet-partner state $M_{\chi_{c2}^{\prime}}=3929\pm 5$~MeV~\cite{uehara_chi3929}.
If the $\chi_{c1}^{\prime}$ mass is $\simeq$3872~MeV, the 
$\chi_{c2}^{\prime}$-$\chi_{c1}^{\prime}$ mass splitting would be
$\simeq 57$~MeV, and higher than the $\chi_{c2}$-$\chi_{c1}$ mass splitting
of $45.5\pm 1.1 $~MeV.  In potential models this splitting decreases with increasing
radial quantum numbers~\cite{splitting}.  Second, the decay
$\chi_{c1}^{\prime}\rt \gamma \psip$
is a favored $E1$ transition and expected to be more than an order-of-magnitude
stronger than ``hindered'' $E1$ transition $\chi_{c1}^{\prime}\rt\gamma\jp$~\cite{NR}.
The Belle experiment recently
reported a 90\% CL limit on  $\Gamma_{X(3872)\rt \gamma \psip}$ that
is less than  $2.1\times\Gamma_{X(3872)\rt \gamma \jp }$~\cite{belle_gammajp}
and in contradiction with potential model expectations for the
$X(3872)=\chi_{c1}^{\prime}$ assignment.  Third, the transition
$\chi_{c1}^{\prime}\rt\pipi\jp$, the $X(3872)$ discovery mode, violates
isospin and is expected to be strongly suppressed.   

Two features of the $X(3872)$ that have attracted considerable attention are its
narrow natural width, $\Gamma_{X(3872)}<1.2$~MeV at the 90\% CL~\cite{belle_jpc},
and its mass, for which (my) world average value is $M_{X(3872)}=3871.67\pm 0.17$~MeV, 
which is equal, to about a part in $\sim 10^4$, to the $D^0\bar{D^{*0}}$ mass
threshold:  $m_{D^0}+m_{D^{*0}}= 3871.79\pm 0.30$~MeV.\cite{pdg} The close proximity
of the $M_{X(3872)}$ to the $D^0\bar{D^{*0}}$ threshold has led to speculation that
the $X(3872)$  is a molecule-like $D^0$-$\bar{D^{*0}}$ bound state held together
by nuclear-like $\pi$- and $\omega$-meson exchange forces~\cite{molecule}.

\subsection{The $Y(4260)$}
\noindent
The $Y(4260)$ was seen by BaBar as a peak in the $M(\pipi\jp)$
distribution in the initial-state-radiation (ISR) process
$\ee\rt\gamma_{ISR}\pipi\jp$~\cite{babar_y4260}, an observation
that was confirmed by CLEO and Belle~\cite{belle_y4260}.  Since it
is produced via the ISR process, its $J^{PC}$ must be $1^{--}$. In
contrast to the $X(3872)$, the peak is relatively wide; the
weighted average of the BaBar and Belle peak width measurements is 
$\Gamma_{Y(4260)}=99\pm 17$~MeV.   

A striking feature of the $Y(4260)$ is that its peak mass is
not near that of any of the established $1^{--}$ charmonium states. Moreover,
since all $1^{--}$ charmonium states with mass below $4.45$~MeV have
been identified, the $Y(4260)$ cannot be a standard $\ccbar$ meson.
Moreover, it does not seem to have a strong coupling to open-charm
mesons; measurements of $\ee$ annihilation into charmed mesons in the
vicinity of $\sqrt{s}\sim 4260$~MeV show indications of a dip in the
total cross section at the location of the $Y(4260)$
peak~\cite{bes2_rscan}.  This motivated a detailed analysis~\cite{xhmo}
that established a lower limit on the partial width
$\Gamma_{Y(4260)\rt\pipi\jp}$ that is greater than 1~MeV,
which is huge for charmonium.  The Belle group did a 
comprehensive search for $Y(4260)$ decays to all possible final
states containing open charmed meson pairs and found no sign of
a $Y(4260)$ signal in any of them~\cite{galina}. Thus, it seems likely that
the  $\Gamma_{Y(4260)\rt\pipi\jp}$ is
substantially greater than 1~MeV.  If this is the case, it 
would be a strong indication that some new, previously unanticipated,
mechanism is involved.

Subsequent studies of the $\ee\rt\gamma_{ISR}\pipi\psip $ ISR
process led to discoveries of states with similar characteristics
decaying to the $\pipi\psip$ final state: the $Y(4360)$ by
BaBar~\cite{babar_y4360}, and the $Y(4660)$ by
Belle~\cite{belle_y4660}. There is no evidence for open-charmed
meson decays for either of these states.  Moreover, there is
no sign of them in the $\pipi\jp$ spectrum and is there no 
evidence for $Y(4260)\rt \pipi\psip $.

\section{Searches in the $b$-quark sector}
\noindent
The existence of the $Y(4260)$ and other $1^{--}$ hidden charm states
with large partial widths to $\pipi\jp$ and $\pipi\psip $
led to speculation that there may be counterparts in the $b$-quark
sector~\cite{wshou}.  This prompted a Belle measurement of the
partial widths for $\Upsilon(5S)\rt\pipi\Upsilon(nS)$
($n=1,2,3$). The expected branching fraction for these
decays~\cite{wshou} is $\sim 10^{-5}$
and, with the data sample that was available at the time, the
expectation was that no signal would be seen.
 (The measured 
branching fractions for the nearby $\Upsilon(4S)$ to
$\pipi\Upsilon(nS)$ are less than $10^{-4}$~\cite{pdg}.)  Rather remarkably,
very strong signals were observed for all three decays modes, with
branching fractions of nearly one percent --- more than
two-orders-of-magnitude times expectations~\cite{kfchen_1}.
In an attempt to determine whether or
not the anomalous events were coming from decays of $\Upsilon(5S)$
or from some other, $b$-quark sector equivalent of the $Y(4260)$ lurking nearby, Belle
did a cross section scan of $\ee \rt $~hadrons and 
$\ee\rt\pipi\Upsilon(nS)$~\cite{kfchen_2} . This scan showed
some indication that the $\ee\rt\pipi\Upsilon(nS)$ yield peaks
at a mass distinct from that for $\Upsilon(5S)\rt$~hadrons
but with limited statistical significance
($10888\pm 3$~MeV for the three $\pipi\Upsilon(nS)$
channels {\it vs.} $10879\pm 3$~MeV for inclusive hadrons).

\subsection{Study of inclusive $\Upsilon(5S)\rt\pipi$ plus {\rm anything}}
\begin{wrapfigure}{l}{6.6cm}   
       \centerline{\includegraphics[width=6.6 cm,height=5.5 cm]
                                   {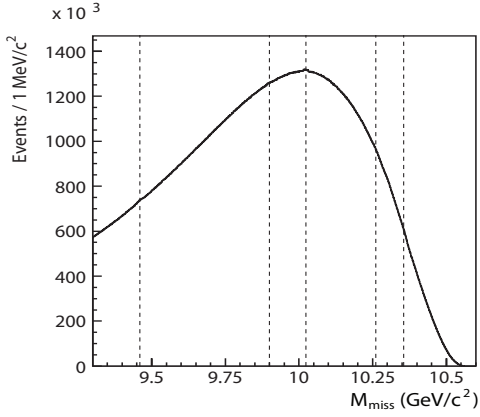}}
      \caption{The mass of the system recoiling against the
      $\pi^+$ and $\pi^-$ in inclusive $\Upsilon(5S)\rt\pipi$~X
      decays.  The dashed lines indicate the  positions
      of the $\Upsilon(1S)$, $h_b(1P)$, $\Upsilon(2S)$, $h_b(2P)$ and
      $\Upsilon(3S)$.}      
      \label{fig:ups5S_incl}
    \end{wrapfigure}

\noindent
Motivated by the curious phenomena described in the preceding section, 
Belle made a study of the inclusive process
$\Upsilon(5S)\rt\pipi$~plus~{\it anything}~\cite{belle_hb}. 
Figure~\ref{fig:ups5S_incl} shows the missing mass recoiling against
every $\pipi$ pair in events in a 121~fb$^{-1}$ data sample collected at an $\ee$
c.m. energy in the vicinity of the $\Upsilon(5S)$ resonance.
In this plot there are a huge number of entries, on the order of a million
in each of the 1~MeV bins; the relative statistical error on each point is of order $0.1\%$.
The distribution is fitted piecewise to a polynomial background shape
plus signal peaks for all of the bottomonim states (and reflections)
that are expected to be produced via this process.

\begin{figure}[b]
       \centerline{\includegraphics[width=9 cm, height=5.5 cm]
                                   {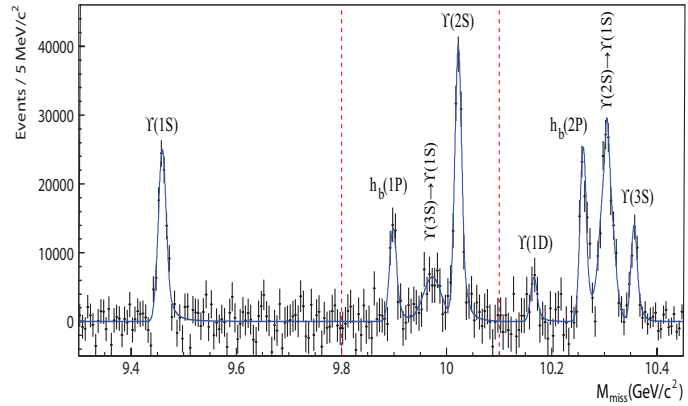}}
   \caption{The background-subtracted recoil mass distribution
    with the signal component from the fit superimposed.  The
    vertical lines indicate the boundaries used for the piecewise
    fit.}
   \label{fig:h_bnP}
   \end{figure}

Figure~\ref{fig:h_bnP} shows the results of the fit
with the background component subtracted.  There, in addition
to peaks corresponding to $\pipi\Upsilon(nS)$ ($n=1,2,3$) and
reflections from the ISR processes $\ee\rt\gamma_{ISR}\Upsilon(mS)$,
$\Upsilon(mS)\rt\pipi\Upsilon(1S)$ ($m=1, 2$), are distinct
signals for $\Upsilon(5S)\rt\pipi h_b(1P)$ and $\pipi h_b(2P)$
with $ 5.5\sigma$ and $11.2\sigma$ significance, respectively, and a hint 
of $\Upsilon(5S)\rt\pipi\Upsilon(1D)$.
This is the first observation of the $h_b(1P)$ and $h_b(2P)$ bottomonium
states.  The prominent $h_b$ signals -- similar in strength to
the $\Upsilon(nS)$ signals -- are somewhat surprising because the
$\Upsilon(5S)\rt\pipi h_b$ process requires a $b$-quark spin-flip
and is expected to be supressed.

\subsection{$M(\pi^{\pm}h_b(mP))$ distributions}
\begin{figure}[t]
\begin{minipage}{70mm}
\centering
 \includegraphics[height=0.6\textwidth,width=0.8\textwidth]{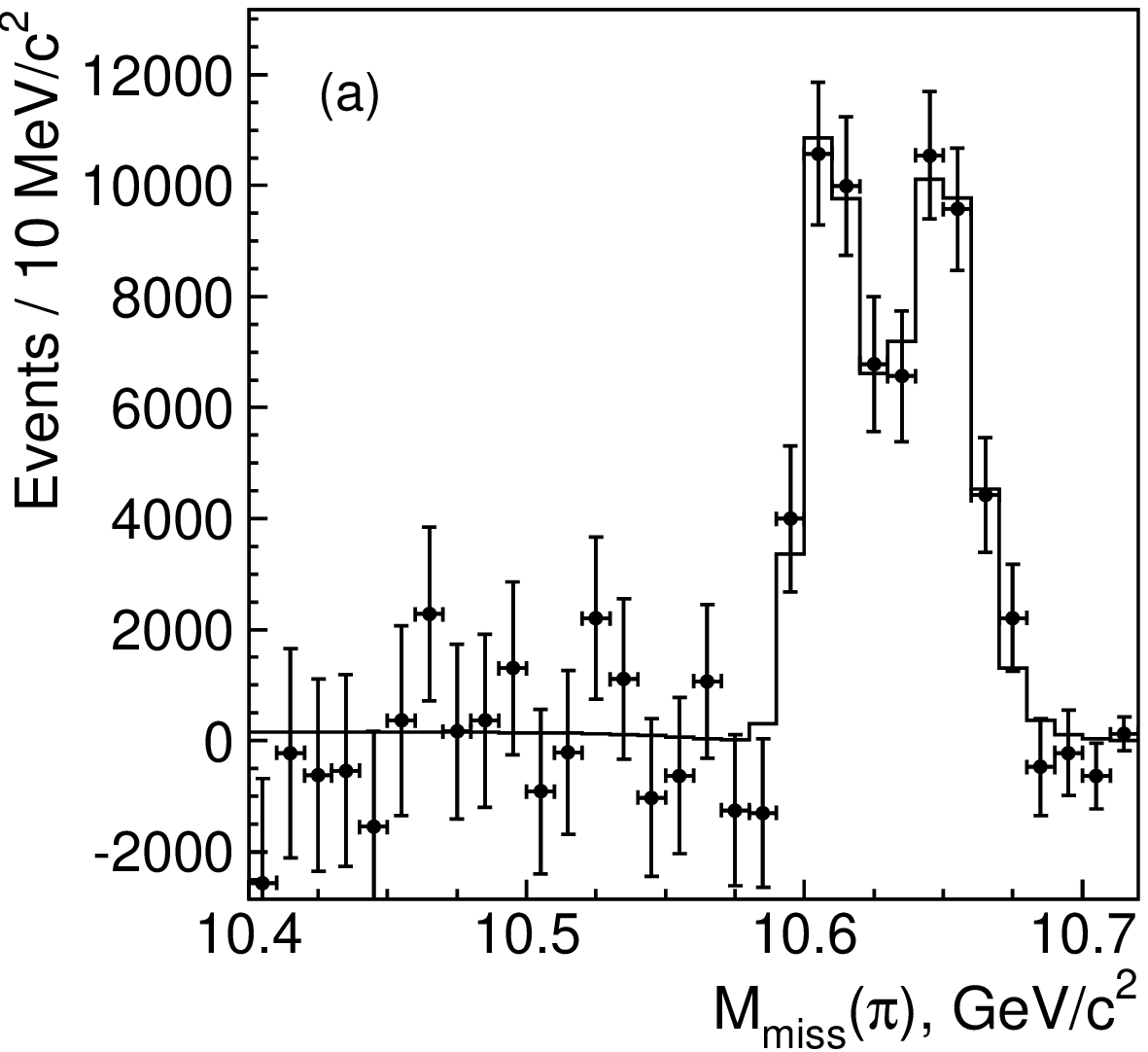}
\end{minipage}
\hspace{\fill}
\begin{minipage}{70mm}
\centering
  \includegraphics[height=0.6\textwidth,width=0.8\textwidth]{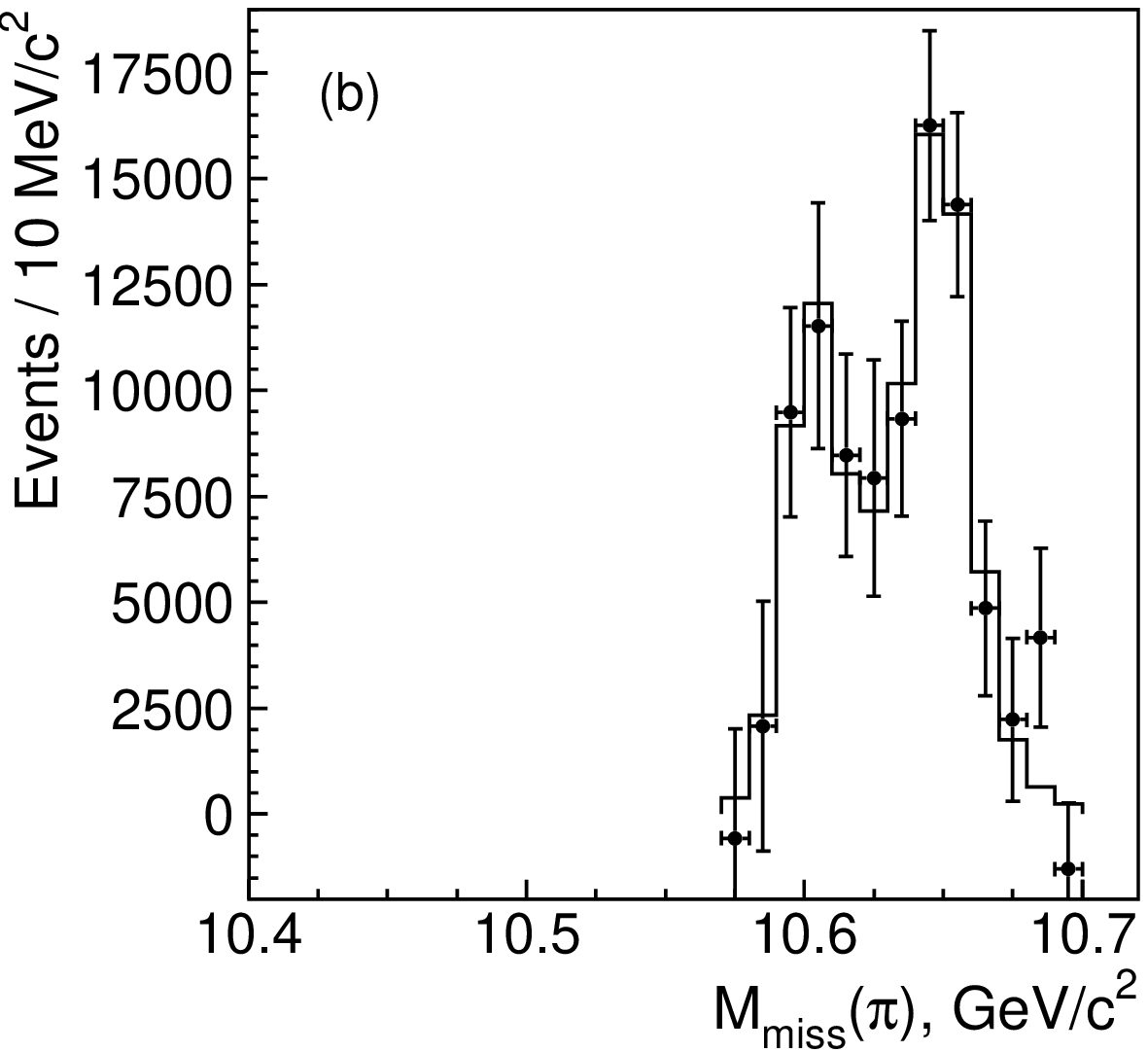}
\end{minipage}
        \caption{(a) $h_b(1P)$ and (b) $h_b(2P)$ yields 
        {\it vs.} $M_{\rm miss}(\pi)$.  The histograms are the fit results.}
        \label{fig:h_bpi}
        \end{figure}

\noindent
The huge number of events in the $h_b(1P)$ and $h_b(2P)$ signal
peaks in Fig.~\ref{fig:h_bnP} ($\simeq50$\ K and $\simeq 84$\ K
events, respectively) permitted an investigation of the
resonant substructure in $\Upsilon(5S)\rt \pipi h_b(mP)$ decays.\cite{bondar_zb}
Figure~\ref{fig:h_bpi}(a) shows the $h_b(1P)$ yield
determined from fits to the $\pipi$ recoil mass spectrum for
different values of $\pi^{\pm }h_b(1P)$ mass, determined from $\pipi$ missing mass
measurements of the $h_b$ signals in bins of the mass recoiling against one of
the pions. Figure~\ref{fig:h_bpi}(b) shows the corresponding
$\pi^{\pm}h_b(2P)$ mass distribution.

As is evident from the figures, the $h_b(1P)$ and $h_b(2P)$ signals
are entirely due to two structures in $M(\pi^{\pm} h_b(mP))$, one with
peak mass near $10610$~MeV and the other
with peak mass near $10650$~MeV.  In the
following, these structures are referred to as the
$Z_b(10610)$ and $Z_b(10650)$, respectively. The
histograms in each figure show the results of fits to the mass
spectra using two Breit Wigner (BW) amplitudes to represent the $Z_b$
peaks plus a phase-space component.  The fitted results for the BW
parameters for the two $Z_b$ peaks, which are consistent with being the
same for both decay channels, are listed below in Table~\ref{tbl:fits}. 
For both spectra, the
fitted strengths of the phase space term are consistent with being zero.

\subsection{$M(\pi^{\pm}\Upsilon(nS))$ distributions in
$\Upsilon(5S)\rt\pipi\Upsilon(nS)$ decays}

    \begin{wrapfigure}{r}{6.6cm}   
       \centerline{\includegraphics[width=6.0 cm,height=5.0 cm]
                                   {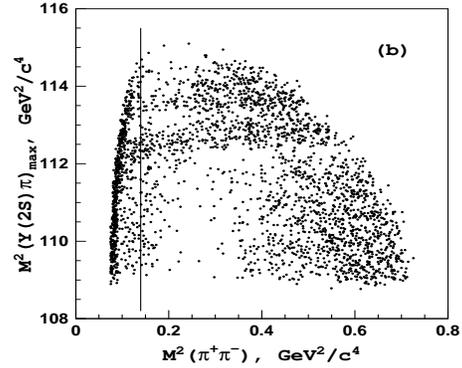}}
      \caption{$M^2(\Upsilon(2S)\pi)$ {\it vs.} 
       $M^(\pipi)$ Dalitz plot for
       $\Upsilon(5S)\rt\pipi\Upsilon(2S)$ decays.}      
      \label{fig:ups2S-pipi-dp}
    \end{wrapfigure}

\noindent
Belle also made an investigation of possible resonant
substructure in fully reconstructed 
$\Upsilon(5S)\rt\pipi\Upsilon(nS)$ decays ($n=1,2,3$)~\cite{bondar_zb}.
Figure~\ref{fig:ups2S-pipi-dp} shows the
$M^2(\Upsilon(2S)\pi)$ (vertical) {\it vs.} 
$M^(\pipi)$ (horizontal) Dalitz plot for
$\Upsilon(5S)\rt\pipi\Upsilon(2S)$ decays.
Here, to avoid double counting, only the
highest mass $\Upsilon(2S)\pi$ combination
is plotted.  In the figure there is
a sharp vertical band at small $\pipi$ masses
caused by background from converted photons,
and two distinct horizontal
clusters near $M^2(\Upsilon(2S)\pi)=112.6$~GeV$^2$
and $113.3$~GeV$^2$, near the locations
expected for the $Z_b(10610)$ and $Z_b(10650)$.  
The $\pipi\Upsilon(1S)$ and $\pipi\Upsilon(3S)$
Dalitz plots show similar structures.

\begin{figure}[h]
\begin{minipage}{70mm}
\centering
 \includegraphics[height=0.6\textwidth,width=0.8\textwidth]{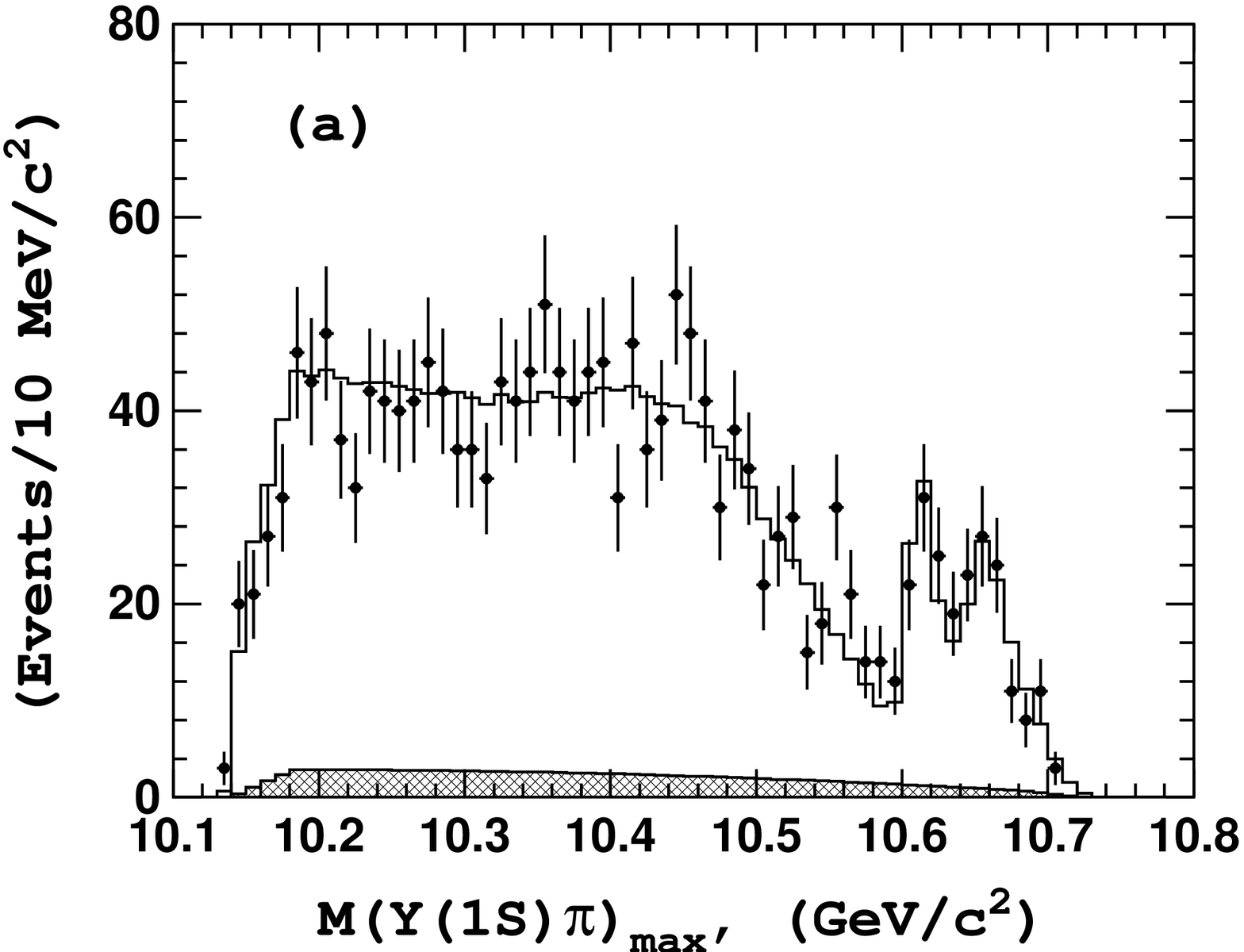}
\end{minipage}
\hspace{\fill}
\begin{minipage}{70mm}
\centering
  \includegraphics[height=0.6\textwidth,width=0.8\textwidth]{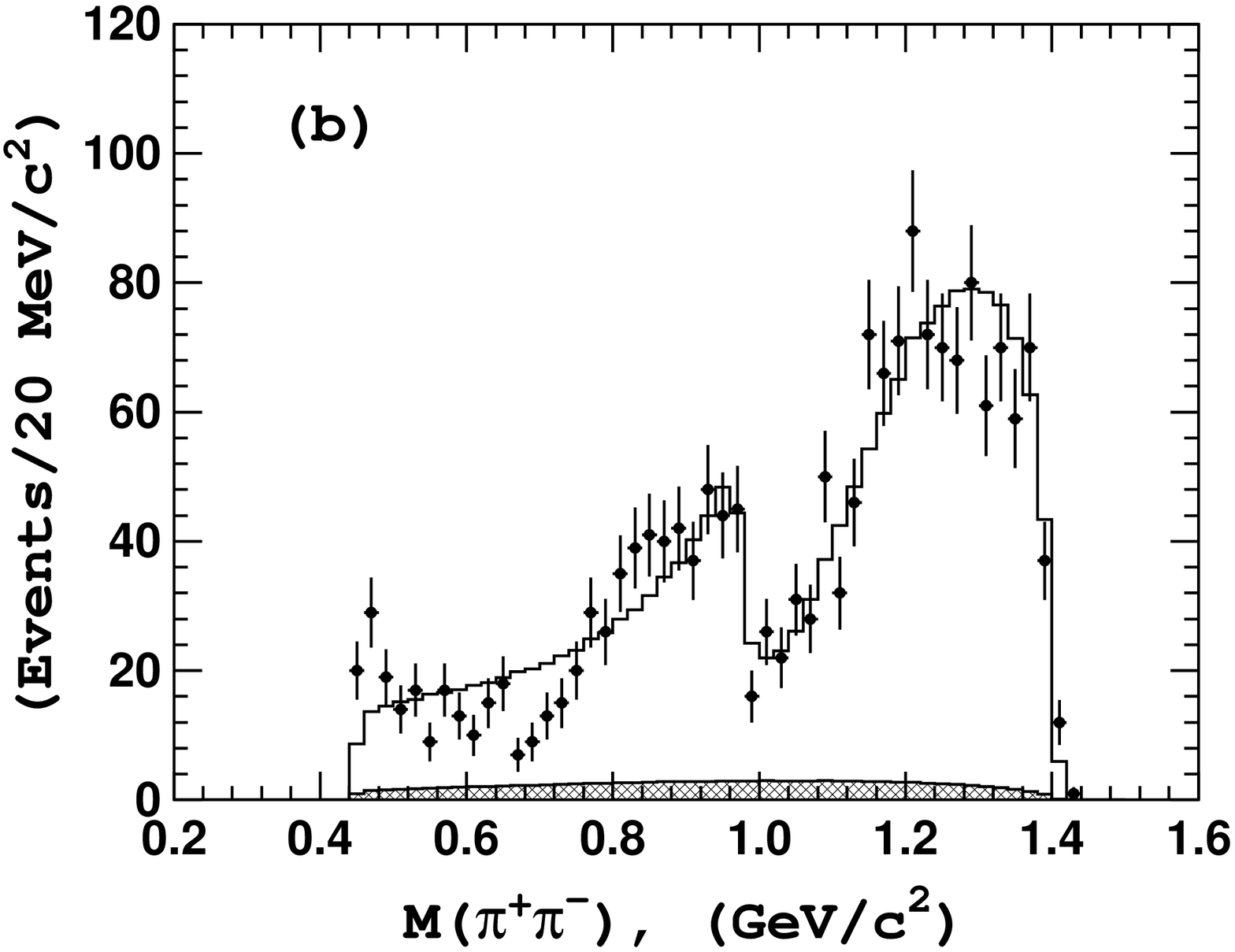}
\end{minipage}

\begin{minipage}{70mm}
\centering
 \includegraphics[height=0.6\textwidth,width=0.8\textwidth]{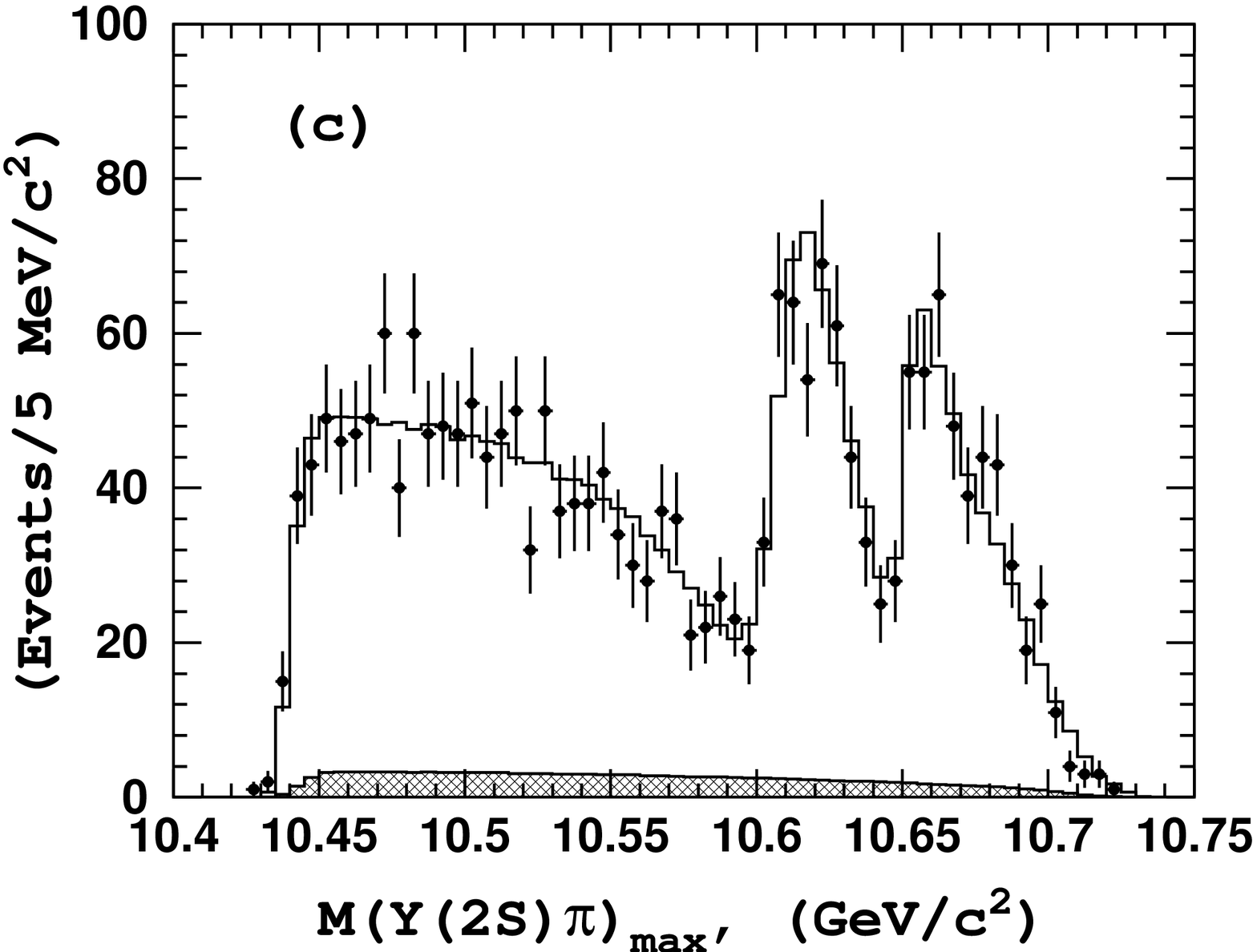}
\end{minipage}
\hspace{\fill}
\begin{minipage}{70mm}
\centering
  \includegraphics[height=0.6\textwidth,width=0.8\textwidth]{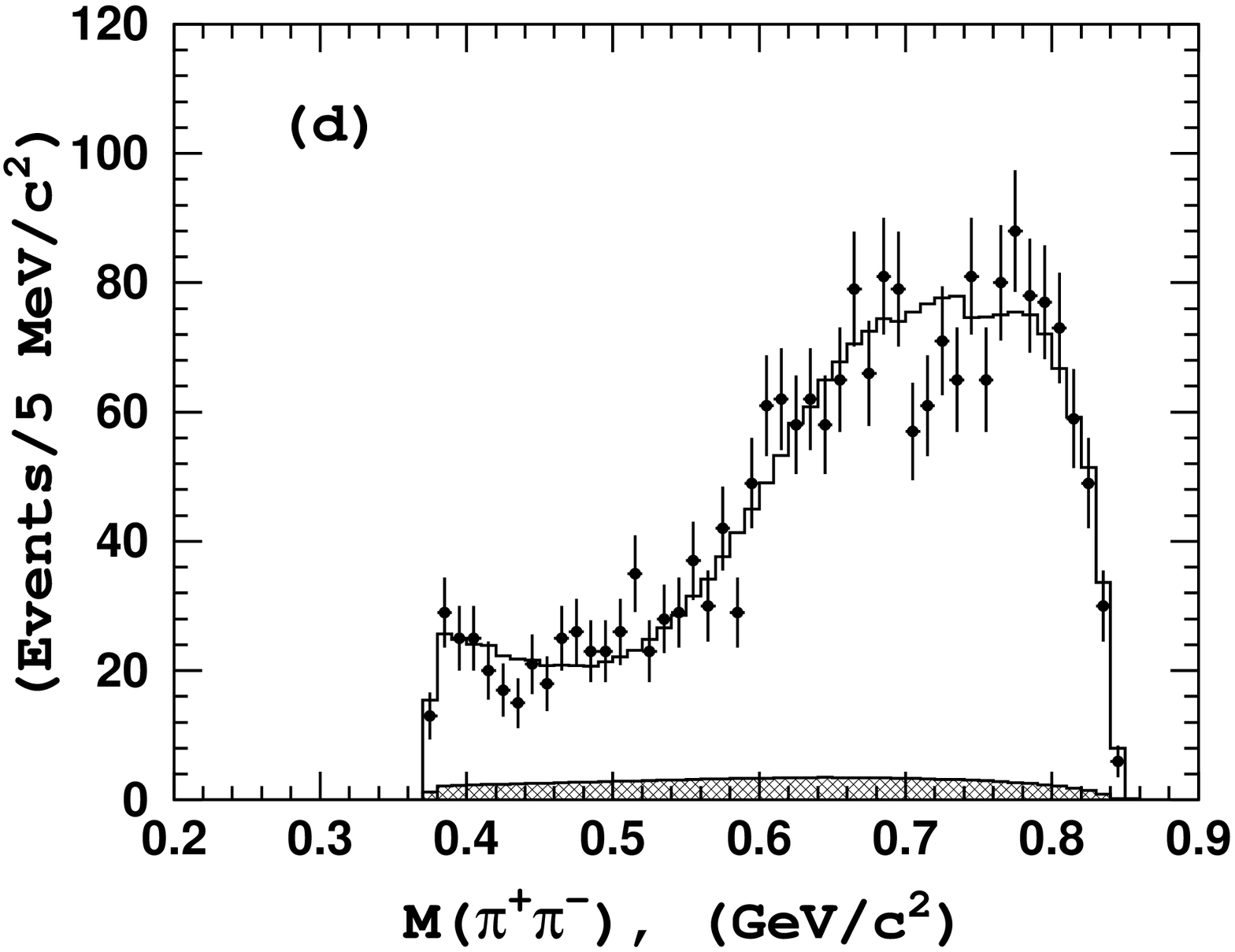}
\end{minipage}

\begin{minipage}{70mm}
\centering
 \includegraphics[height=0.6\textwidth,width=0.8\textwidth]{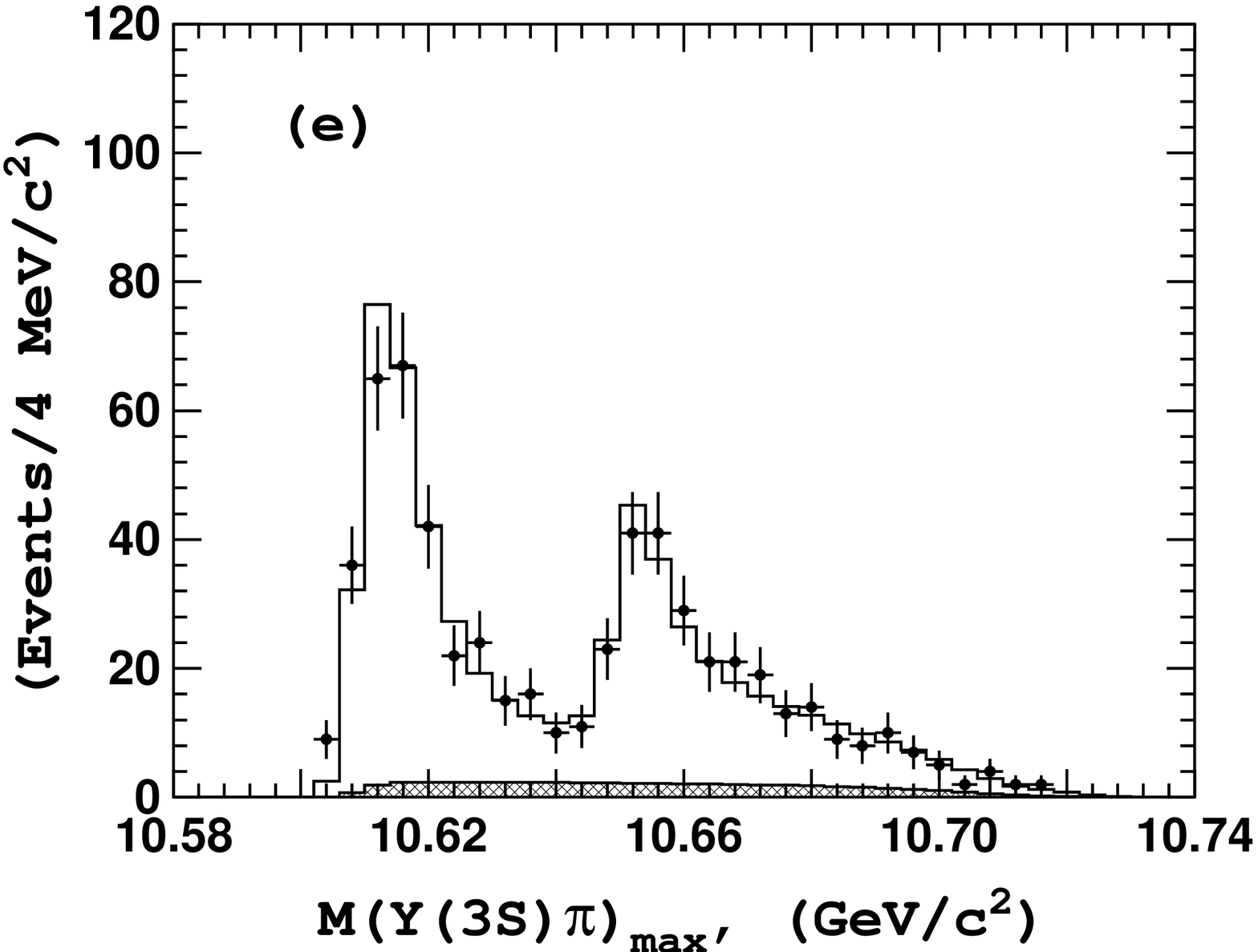}
\end{minipage}
\hspace{\fill}
\begin{minipage}{70mm}
\centering
  \includegraphics[height=0.6\textwidth,width=0.8\textwidth]{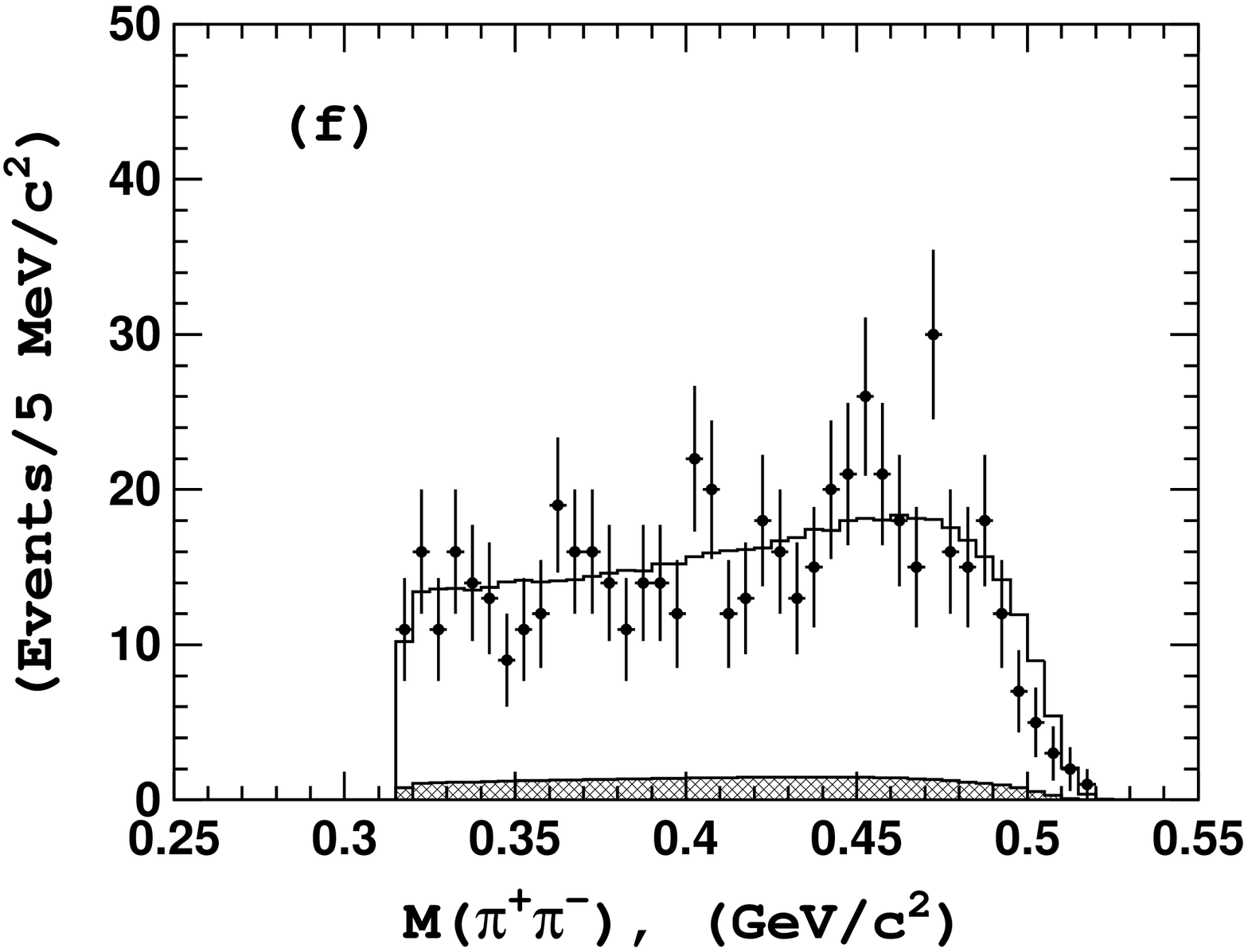}
\end{minipage}
    \label{fig:dp-fits}
    \caption{$M(\Upsilon(nS)\pi)$ and $M(\pipi)$ projections with fit results superimposed
     for the $\Upsilon(1S)$ (a,b), $\Upsilon(2S)$ (c,d) and $\Upsilon(3S)$ signals. The hatched
     histograms are sideband-determined backgrounds.}

        \end{figure}

The Dalitz plots are fitted with a model that includes BW
amplitudes to represent the two $Z_b$ states,  terms that
account for possible contributions in the $\pipi$ system
from the $f_0(980)$ and $f_2(1270)$ resonances, and a
non-resonant amplitude using a form suggested by
Voloshin~\cite{voloshin}. The regions of the Dalitz plots
contaminated by photon conversion backgound ({\it i.e.} to the left
of the vertical line in Fig.~\ref{fig:ups2S-pipi-dp}) are
excluded from the fits.  
$M(\Upsilon(nS)\pi)$ and $M(\pipi)$
projections with the results of the fits superimposed are
shown in Fig.~\ref{fig:dp-fits} and included in Table~\ref{tbl:fits}.  
The $Z_b(10610)$ and $Z_b(10650)$ mass and width measurements from the
five different channels agree within their errors.   The averages of the
five mass and width measurements for the $Z_b(10610)$ are  
$M=10607.2\pm 2.0$~MeV and $\Gamma= 18.4\pm 2.4$~MeV; for the 
$Z_b(10650)$, the averages are 
$M=10652.2\pm 1.5$~MeV and $\Gamma= 11.5\pm 2.2$~MeV.  These
are very near the $m_B + M_{B^*}=10604.3\pm 0.6$~MeV and 
$2m_{B^*}=10650.2\pm 1.0$~MeV\cite{pdg} mass thresholds, respectively,
which is suggestive of virtual molecule-like structures~\cite{bondar},
although other interpretations have been proposed~\cite{other_zb}.

\begin{table}[h!]
  \caption{Results for the $Z_b(10610)$ and $Z_b(10650)$ parameters
           obtained from $\Upsilon(5S)\rt\pipi\Upsilon(nS)$ ($n=1,2,3$) and 
           $\Upsilon(5S)\rt h_b(mP)\pipi$ ($m=1,2$) analyses. }
  \medskip
  \label{tbl:fits}
\centering
  \begin{tabular}{lccccc} \hline \hline
Final state & $\Upsilon(1S)\pi^+\pi^-$      &
              $\Upsilon(2S)\pi^+\pi^-$      &
              $\Upsilon(3S)\pi^+\pi^-$      &
              $h_b(1P)\pi^+\pi^-$           &
              $h_b(2P)\pi^+\pi^-$          \\ \hline
           $M[Z_b(10610)]$, MeV            &
           $10611\pm4\pm3$                  &
           $10609\pm2\pm3$                  &
           $10608\pm2\pm3$                  &
           $10605\pm2^{+3}_{-1}$            &
           $10599{^{+6+5}_{-3-4}}$
 \\
           $\Gamma[Z_b(10610)]$, MeV  &
           $22.3\pm7.7^{+3.0}_{-4.0}$       &
           $24.2\pm3.1^{+2.0}_{-3.0}$       &
           $17.6\pm3.0\pm3.0$               &
           $11.4\,^{+4.5+2.1}_{-3.9-1.2}$   &
           $13\,^{+10+9}_{-8-7}$
 \\
           $M[Z_b(10650)]$, MeV             &
           $10657\pm6\pm3$                  &
           $10651\pm2\pm3$                  &
           $10652\pm1\pm2$                  &
           $10654\pm3\,{^{+1}_{-2}}$        &
           $10651{^{+2+3}_{-3-2}}$
 \\
           $\Gamma[Z_b(10650)]$, MeV  &
           $16.3\pm9.8^{+6.0}_{-2.0}$~      &
           $13.3\pm3.3^{+4.0}_{-3.0}$       &
           $8.4\pm2.0\pm2.0$                &  
           $20.9\,^{+5.4+2.1}_{-4.7-5.7}$   & 
           $19\pm7\,^{+11}_{-7}$ 
\\
\hline \hline
\end{tabular}
\end{table}

\subsection{The transitions $h_b(1P,2P)\rt \gamma\eta_b(1S,2S)$ and the discovery of the $\eta_b(2S)$}
\noindent
In studies of bottomonium physics, the $\Upsilon(1S)$-$\eta_b(1S)$ mass difference has special importance
since this determines the scale of the spin-spin hyperfine interaction term in the $\bbar$ potential.
This is accessible to Lattice QCD calculations, which give values that from
$\Delta_{\rm hfs}(1S)=47$~MeV to $59$~MeV~\cite{meinel}. 
The $\eta_b(1S)$ was discovered by the BaBar collaboration in the $M1$ radiative
process $\Upsilon(3S)\rt \gamma \eta_b(1S)$~\cite{babar_etab}.  BaBar produced the first measurement
of the splitting to be $\Delta_{\rm hfs}(1S)=71.4\pm 4.1$~MeV, which is outside the theoretical range.  This
measurement was an experimental {\it tour-de-force} because the  $\Upsilon(3S)\rt\gamma\eta_b$ 
environment is very difficult, with a weak signal and substantial backgrounds that make the
extraction of a precise mass value difficult.

    \begin{wrapfigure}{r}{5.6cm}   
       \centerline{\includegraphics[width=5.0 cm,height=12.0 cm]
                                   {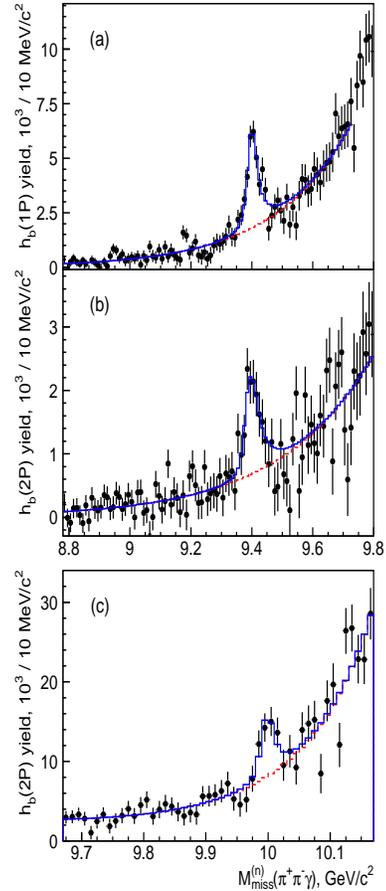}}
      \caption{{\bf a)} The $h_b(1P)$ yield {\it vs.} $M_{\rm miss}(\pipi\gamma)$
       and {\bf b)} the $h_b(2P)$ yield {\it vs.} $M_{\rm miss}(\pipi\gamma)$
       inn the $\eta_b(1S)$ mass region. {\bf c)} The $h_b(2P)$ yield {\it vs.} $M_{\rm miss}(\pipi\gamma)$
       inn the $\eta_b(2S)$ mass region.}
      \label{fig:hb_2_etab}
    \end{wrapfigure}

The Belle observation of strong signals for $h_b(1s)$
and $h_b(2S)$ in inclusive $\Upsilon(5S)\rt \pipi X$ decays, provides another way to access
the $\eta_b(1S)$, and that is via the $E1$ transitions, $h_b(1P,2P)\rt \gamma\eta_b(1S)$.  
Figures~\ref{fig:hb_2_etab}(a) and (b) show the $h_b(1S)$ and $h_b(2P)$ signal yields determined from
fitting the $\pipi$-recoil mass spectra, but this time in bins
of $\pipi\gamma$ missing mass. In this measurement,  $\gamma$s that are
not associated with a $\pi^0\rt\gamma\gamma$ decay are combined with the $\pipi$ pairs
to determine $M_{\rm miss}(\pipi\gamma)$, which is plotted on the horizontal axis~\cite{belle_etab}.  
Distinct peaks near $9.4$~GeV corresponding to the $\eta_b(1S)$
are evident in both distributions. These data are used to determine the hyperfine splitting
$\Delta_{\rm hfs}(1S)=57.9\pm 2.3$~MeV and total width $\Gamma_{\eta_b(1S)}=10.8\pm 5.8$~MeV.  This
measurement of $\Delta_{\rm hfs}(1S)$ has improved precision and 
has a central value that is about $2.9\sigma$ higher
than the BaBar measurement and within the range of theoretical expectations.    

Figures~\ref{fig:hb_2_etab}(c) shows the $h_b(2P)$ signal yields determined from
fitting the $\pipi$ recoil spectrum in $\pipi\gamma$ bins in the mass range expected for
the $\eta_b(2S)$, where a prominent peak can be seen near $10$~GeV.  Belle identifies this
as the first observation of the $\eta_b(2P)$ and measures $\Delta_{\rm hfs}(2S)=24.3\pm 4.3$~MeV.

As mentioned above, the LQCD calculations of $\Delta_{\rm hfs}(nS)$ produce a range of values that
reflect the different approximations that are necessary for managable lattice calculations.
On the other hand, in ratios of the splittings between different radial states, many of
these uncertainties cancel.  Thus, at least for the time being, measurements of these ratios
present the strongest challenges for theory.  The Belle measurement of the ratio
$\Delta_{\rm hfs}(2S)/\Delta_{\rm hfs}(1S) = 0.42\pm0.08$ is in agreement with a LQCD 
prediction of $0.40\pm 0.06$~\cite{meinel}.

\section{Search for the $H$-dibaryon in $\yones$ and $\ytwos$ decays.}
\noindent
As mentioned in the intoduction above,
recent theoretical results motivate searches for the $H$
with mass near the $M_H = 2m_{\Lambda}$ threshold. This mass
region is especially interesting, because very general theoretical
arguements ensure  that for masses 
approaching the $2m_{\lm}$ threshold from below, the
$H$ would behave more and more like a $\lm\lm$
analog of the deuteron, and for masses approaching $2m_{\lm}$ from
above, the $H$ would look more and more like a virtual dineutron resonance,
 independently of its dynamical origin~\cite{braaten}. If its mass is below 
$2m_{\lm}$, the $H$ would predominantly decay via
$\Delta S=+1$ weak transitions to $\lm n$,
$\Sigma^- p$, $\Sigma^0 n$ or $\lmppi$ final states.
If its mass is above $2m_{\lm}$, but below 
$m_{\Xi^0} + m_n~(=2m_{\lm}+ 23.1$~MeV), the $H$ would decay
via strong interactions to $\lmlm$ 100\% of the time.

Decays of narrow $\yns$ ($n=1,2,3$) bottomonium
($b\bar{b}$) resonances are
particularly well suited for searches for deuteron-like
multiquark states with
non-zero strangeness.  The $\yns$ states are flavor-$SU(3)$
singlets and primarily decay via the 
three-gluon annihilation process  ({\it e.g.}, 
($Bf(\yones\rt ggg) = 81.7\pm 0.7$\%~\cite{pdg}). 
The gluons materialize into $\uubar$, $\ddbar$ and $\ssbar$
pairs in roughly equal numbers. 
The high density of quarks and antiquarks in the limited final-state phase
space is conducive to the production of
multi-quark systems, as demonstrated by large branching
fractions for inclusive antideuteron ($\bar{d}$) production:
$Bf(\yones\rt \bar{d}\, X)=
(2.9\pm 0.3)\times 10^{-5}$ and 
$Bf(\ytwos\rt \bar{d}\, X)=
(3.4\pm 0.6)\times 10^{-5}$~\cite{CLEO_dbar}.  An upper limit for
the production of a six-quark $S=-2$ state in $\yns$ decays that is
substantially below that for the six-quark antideuteron would
be strong evidence against its existence.

Belle recently completed a search for
$H$-dibaryon production in the inclusive decay chains 
$\yonetwos \rt H \, X$; $H\rt \lmppi$ and 
$H\rt \lmlm$~\cite{belle_H},
using data samples containing 102 million $\yones$ and
158 million $\ytwos$ decays.  The search strategy
assumed equal $\yones$ and $\ytwos$ branching fractions:
{\it i.e.,} $Bf(\yones\rt H\, X)=Bf(\ytwos\rt H\, X)
\equiv Bf(\yonetwos\rt H\, X)$.

\begin{figure}[htb]
\begin{minipage}{70mm}
\centering
 \includegraphics[height=0.6\textwidth,width=0.8\textwidth]{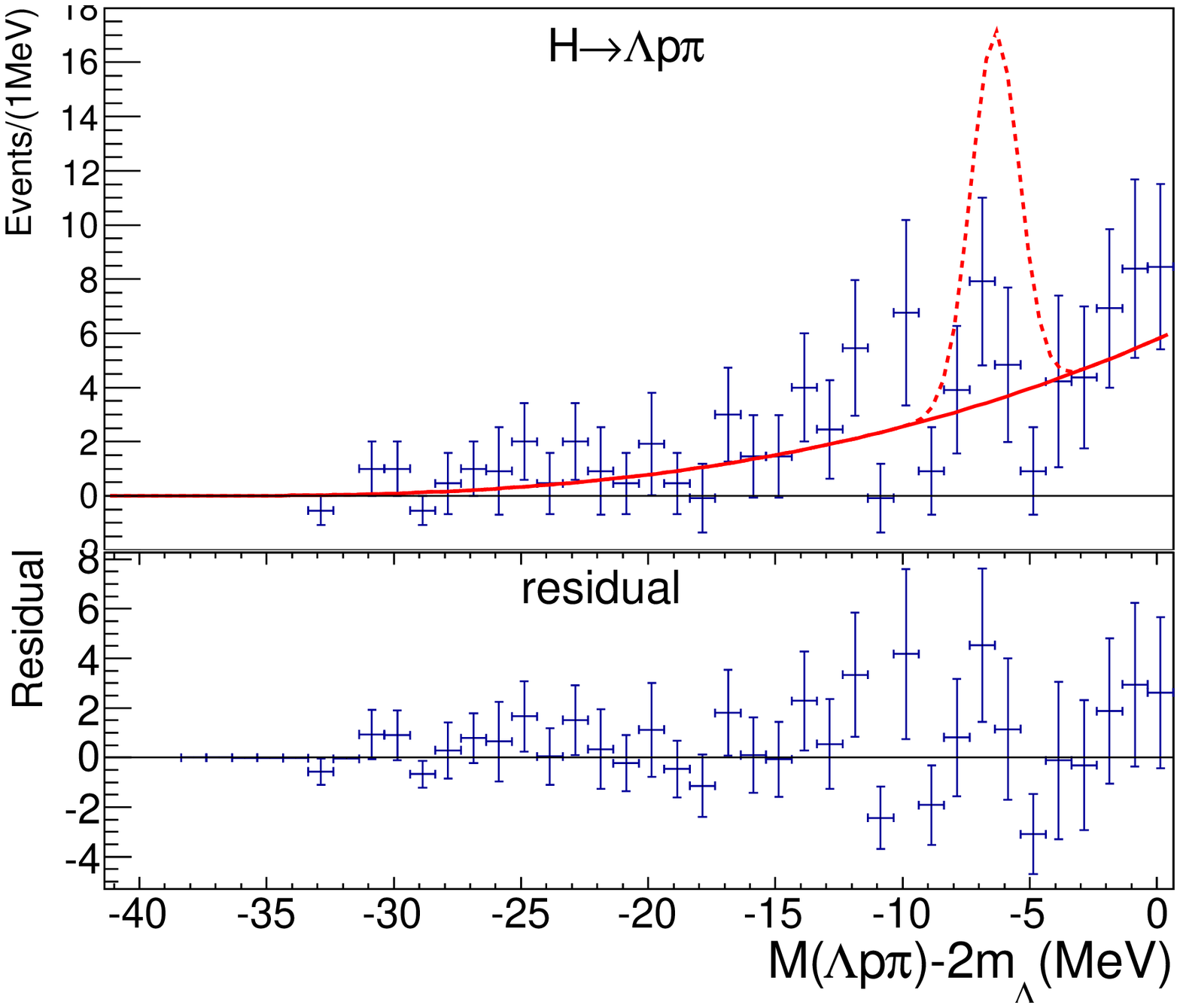}
\end{minipage}
\hspace{\fill}
\begin{minipage}{70mm}
\centering
  \includegraphics[height=0.6\textwidth,width=0.8\textwidth]{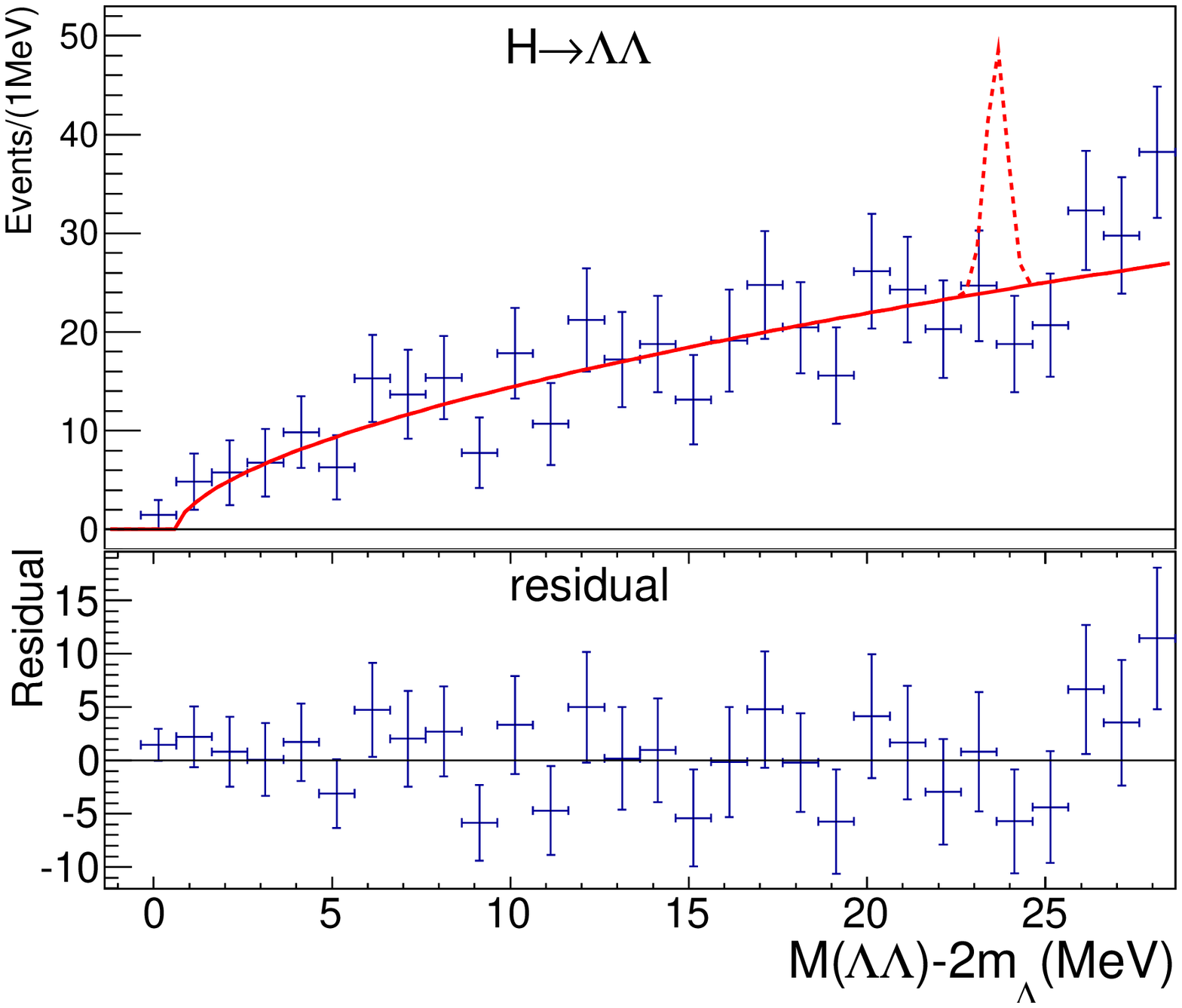}
\end{minipage}

\begin{minipage}{70mm}
\centering
 \includegraphics[height=0.6\textwidth,width=0.8\textwidth]{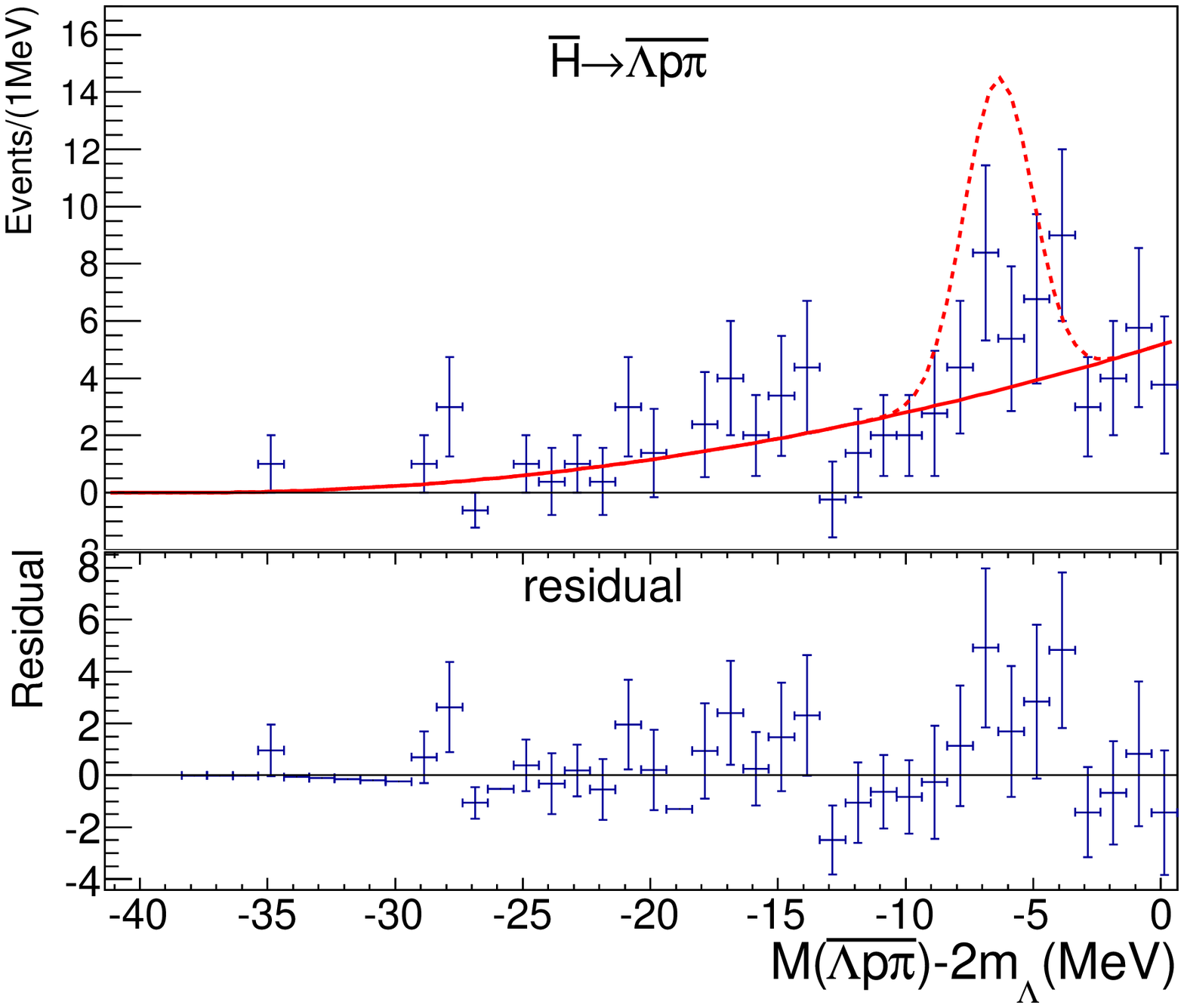}
\end{minipage}
\hspace{\fill}
\begin{minipage}{70mm}
\centering
  \includegraphics[height=0.6\textwidth,width=0.8\textwidth]{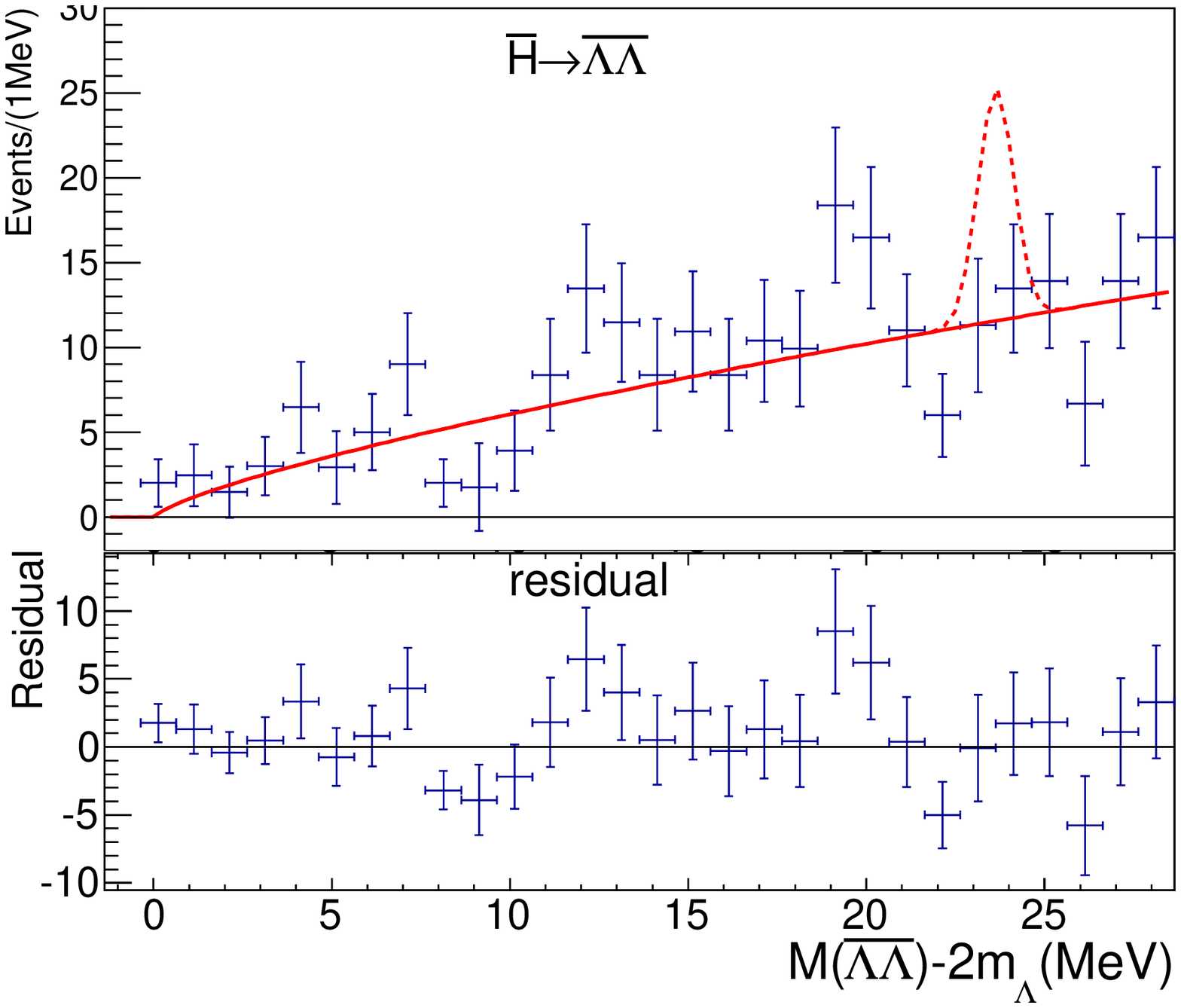}
\end{minipage}
\caption{ 
{\bf Top:}  The continuum-subtracted $M(\lmppi)$ (left) and $M(\lm\lm)$ (right) distributions
with the residuals from a background-only fit shown below. Here the $\yones$ and $\ytwos$ data samples
are combined.  The curve shows the
results of the background-only fit described in the text.  The dashed curve
shows the expected $H$ signal for a $\yonetwos\rt H X$ branching
fraction that is $1/20^{\rm th}$ that for antideuterons.
{\bf Bottom:} The corresponding $M(\lmbpbpi ) $ (left) and $M(\lmb\lmb)$ distributions.
}
\label{fig:data-lppi}
\end{figure}

The resulting continuum-subtracted $M(\lmppi)$ ($M(\lmbpbpi)$)
distribution for the combined $\yones$ and $\ytwos$ samples,
shown in the top (bottom) left-hand panels of Fig.~\ref{fig:data-lppi}, has no evident 
$H\rt\lmppi$ ($\bar{H}\rt\lmbpbpi$) signal.  The curve in the figure is the result of a fit using 
a threshold function to model the background; fit residuals are also shown.
The dashed curves in the figures show the expected $H$ signal for a $\yonetwos\rt H X$ branching
fraction that is $1/20^{\rm th}$ that for antideuterons.

The panels on the right of Fig.~\ref{fig:data-lppi} show the $M(\lm\lm)$ (above) 
and $M(\lmb\lmb)$ (below) distributions for events that satisfy
the selection requirements. Here there is
no sign of a near-threshold enhancement similar to that reported by the
E522 collaboration~\cite{e522} nor any other evident signal for
$H\rt\lm\lm$ ($\bar{H}\rt\lmb\lmb$).  The curve is the result of a
background-only fit using the functional form described above; fit
residuals are also shown.  Expectations for a signal branching fraction
that is $1/20^{\rm th}$ that for the antideuterons is indicated with a dashed curve.

\begin{figure}[htb]
\begin{center}
\includegraphics[height=0.25\textwidth,width=0.5\textwidth]
{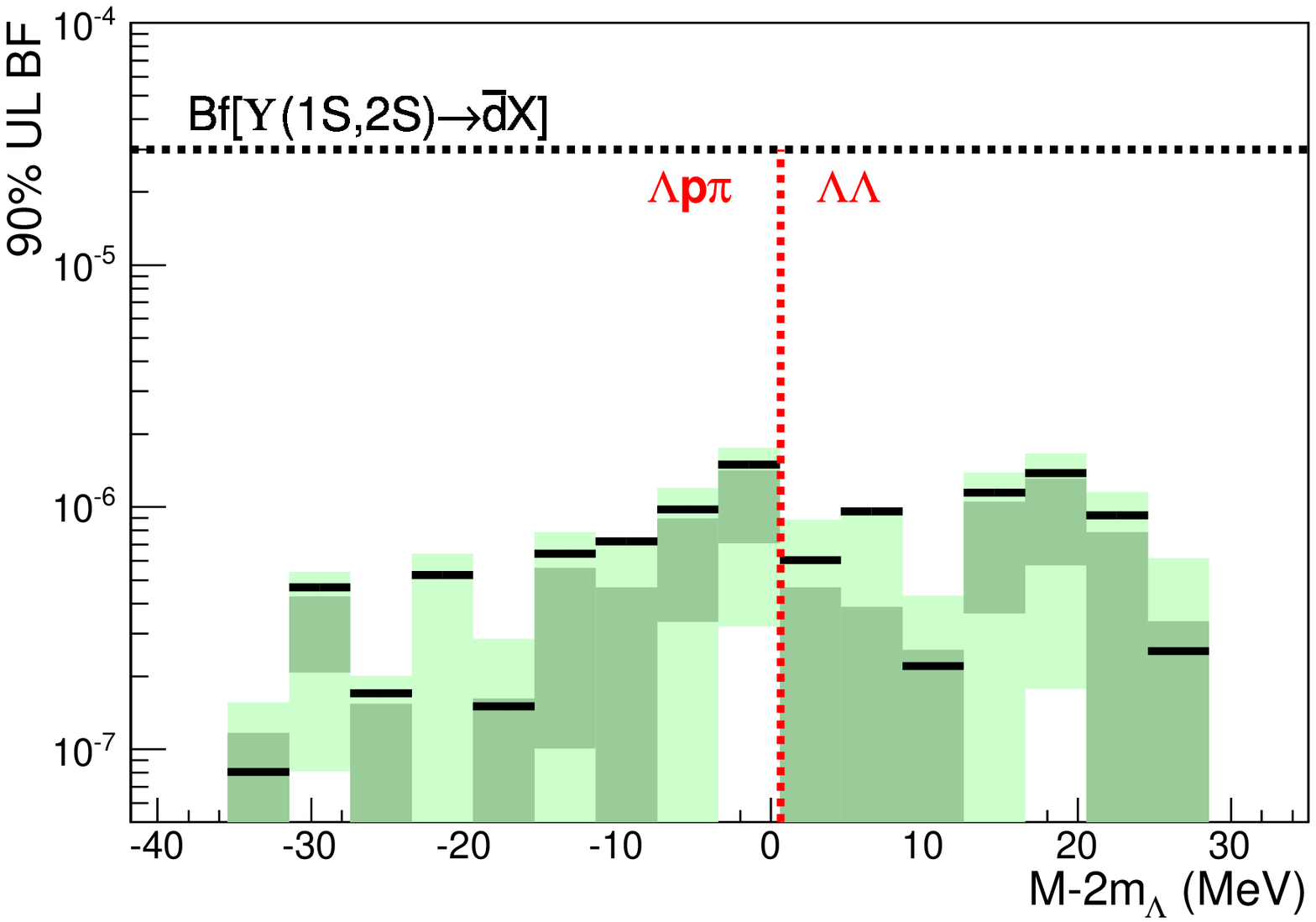}
\end{center}
\caption{ 
Upper limits (at 90\% CL) for 
$Bf(\Upsilon(1S,2S)\rt H~X)\cdot Bf(H \rt f_i)$
for a narrow ($\Gamma=0$) $H$-dibaryon {\it vs.} $M_H-2m_{\lm}$ are shown as
solid horizontal bars.  The one (two) sigma bands are shown as the
darker (lighter) bands.
The vertical dotted line indicates the $M_H=2m_{\lm}$ threshold.  The
limits below (above) the $2m_{\lm}$ threshold are for $f_1= \lmppi$
($f_2=\lm\lm$). The horizontal dotted line indicates the
average PDG value for $Bf(\Upsilon(1S,2S)\rt \bar{d}~X)$.
}
\label{fig:limits}
\end{figure}

In the absence of any sign of an $H$-dibaryon in either the
$\lm p\pi$ or the $\lmlm$ mode, we set the 90\% CL
($M_H - 2m_{\lm}$)-dependent branching fraction upper limits for the $\lmppi$ and
$\lm\lm$ (for $\Gamma=0$) mode shown in
Figure~\ref{fig:limits}.  These limits are all more than an order of magnitude
lower than the average of measured values of
 $Bf(\Upsilon(1,2S)\rt \bar{d}~X)$, shown 
in Fig.~\ref{fig:limits} as a horizontal dotted line.  

These new Belle results are some of the most stringent constraints to date on
the existence of an $H$-dibaryon with mass near the $2m_{\lm}$ threshold~\cite{limits}.
Since  $\Upsilon\rt\ $hadrons decays produce final states that are flavor-$SU(3)$
symmetric, this suggests that if an $H$-dibaryon exists in this mass
range, it must have very different dynamical properties than the deuteron,
or, in the case of $M_H<2m_{\lm}$, a strongly suppressed $H\rt \lmppi$ decay mode.

\section{Comments}
\noindent
After years of theoretical and experimental work, a large assortment of particles,
the $XYZ$ mesons, have been found that can not be accounted for by the standard mesons are
quark-antiquark pairs rule that has been in common practice. There are now
almost twenty candidates, a number that continues to grow rapidly.
Many of these new mesons are close to particle-antiparticle thresholds and look
very much like molecular structures of color-singlet mesons, however others
are far from thresholds, which make molecular assignments less compelling.
One feature of these states are their strong decays to hidden quarkonium states.
In cases where partial width measurements have been possible, the results are
usually much larger than those measured for conventional quarkonium states.
Likewise, decays to open-flavor states seem to be suppressed compared to
those for quarkonium mesons.
 
Few of the observed states were predicted in advance by theorists, while
some predicted states, such as charged partners of the $X(3872)$
have been searched for but not been seen~\cite{belle_jpc}.  Moreover, none
of the new particles make compelling matches to any of the states 
that are predicted by the QCD-motivated models that theorists seem
to really like.  The  experimental limits on pentaquarks and the
$H$-dibaryon keep getting more stringent with no compelling signs
for either of them.  Attempts have been made to attribute some of
the $XYZ$ states to diquark-diantiquark color bound states~\cite{maiani},
however, these models predict that these structures should form
flavor-$SU(3)$ multiplets and, so far at least, no multiplet
partners of the observed states have been found.

This remains very much an experiment-dominated field of research.
 Hopefully as the list of $XYZ$ states expands and the properties of
the established states are better know, some pattern will emerge that
will allow someone to make sense of it all. 

\section{Acknowledgements} 
\noindent
I thank the organizers of this meeting for inviting me to present
these results.  In addition I compliment them on their well organized and
interesting meeting.
This work is supported by the Korean national Research Foundation 
via Grant No. 2011-0029457 and WCU Grant No. R32-10155.


\begin{thebibliography}{99}
\bibitem{conj} In this report the inclusion of charge-conjugate
states is always implied.
\end{thebibliography}
\end{document}